\documentclass[a4paper]{jpconf}
\usepackage{amsthm,graphicx,amsmath,amssymb,amsfonts,framed}
\newcommand{\A}{{\mathcal A}}

\newcommand{\I}{{\mathbb I}}

\newcommand{\M}{{\mathcal M}}
\newcommand{\HH}{{\mathcal H}}
\newcommand{\cinf}{{C^\infty(\M)}}
\begin{document}
\title{Beyond the Standard
  Model with noncommutative geometry, strolling towards quantum gravity}

\author{Pierre Martinetti}

\address{Universit\`a di Trieste, dipartimento di Matematica e Informatica, via Valerio 12/1, I-34127}

\ead{pmartinetti@units.it}

\begin{abstract}
Noncommutative geometry in
its many incarnations appears at the crossroad of various researches in
theoretical and mathematical physics: from models of quantum space-time (with or without breaking of Lorentz
symmetry) to loop gravity and string theory, from early considerations
on UV-divergencies in quantum field theory to recent models of gauge
theories on noncommutative spacetime, from Connes description of the
standard model of elementary particles to recent Pati-Salam like
extensions.  We list
several of these applications, emphasizing also the original point of
view brought by noncommutative geometry on the nature of time. This text serves as an introduction to the volume of proceedings of the parallel
session ``Noncommutative geometry and quantum gravity'',
 as a part of the conference ``Conceptual and technical
challenges in quantum gravity'' organized at the University of Rome
\emph{La Sapienza} in September 2014.
\end{abstract}

\section{Introduction}

Starting with early considerations of Bronstein, Mead, Wheeler,
Pauli, Heisenberg and others (see \cite{Piacitelli:2010uq} for a nice historical view on the
subject), there exists a vast literature dealing with the reasons why putting together
general relativity with quantum mechanics asks for a revision
  of the classical concept of spacetime. A common feature is the
  emergence of noncommutativity in the description of spacetime
  itself, as opposed to quantum mechanics alone where
  noncommutativity lives on the phase space. The appearance of
  spacetime noncommutativity has a wide range of
motivations, from pure mathematics to phenomenological arguments. We shall not try to be
  exhaustive here, and simply point out some
  directions of research that have been through interesting
  developments in the last years. We shall make the
  distinction between \emph{noncommutative spacetimes} intended as
  spaces whose coordinates no longer commute, and \emph{spectral
    geometries} intended as a space whose algebra of functions is
  non-necessarily commutative. 
\medskip

\emph{Noncommutative spacetimes}, as recalled in section 2,
  are obtained as a deformation of a usual space by
trading the (commutative) coordinate functions $x^\mu, x^\nu$ of a
manifold with coordinate operators $q^\mu, q^\nu$ satisfying non-trivial commutation
relations. Besides the seminal quantum spacetime model  of Doplicher, Fredenhagen, Roberts
\cite{Doplicher:1995hc}
treated in L. Tomassini talk
\cite{Tomassini:2015aa}, such spaces are present in many - if not all - approaches to
quantum gravity, including loop quantum gravity, string theory (see
P. Aschieri and R. Szabo text \cite{Aschieri:2015aa}), as
well as in more phenomenology-oriented models like doubly special
relativity, as illustrated by S. Bianco in his presentation \cite{Bianco:2015aa}. 

Noncommutative spacetimes also emerged very early as a
possible solution to ultraviolet divergencies in quantum field theory,
especially in the work of Snyder
\cite{Snyder:1947ix}. Quantum field and gauge theories on
noncommutative spacetimes have thus developed as a theory on their own, independently of any
consideration on quantum gravity. Recent advances are presented in the
texts of T. Juric, S. Meljanac, A, Samasarov one the one side \cite{T.-Juric:2015aa}, 
and  J.-C. Wallet \& A. G\'er\'e for noncommutative gauge theories
\cite{wal-rev} on the other \cite{Gere:2014fk}. 

Nevertheless, revisiting the work of Snyder in the light of
nowadays quantum-gravity problematic offers an intriguing point of
view, which is presented in V. Astuti paper \cite{Astuti:2015aa}.
\medskip

\emph{Spectral geometries} \cite{Connes:1994kx} is the subject of section 3. It consists in a
generalization of Gelfand duality
between locally compact spaces and $C^*$-algebras, so that
to encompass all the aspects of Riemannian geometry
\cite{connesreconstruct} beyond topology. It furnishes a
geometrical interpretation of the Lagrangian of the standard model of
elementary particles \cite{Connes:1996fu,Chamseddine:2007oz}, as well as some possibilities to go
beyond \cite{Chamseddine:2013uq,Devastato:2013fk}. Recent progress on that matter are reported in the
contribution of A. Devastato \cite{Devastato:2015aa}.
Finally F. Besnard \cite{Besnard:2015aa} discusses the extension of this
approach to the pseudo-Riemannian case.

\section{Noncommutative spacetimes}

\subsection{Poincar\'e covariant spacetime}
Spacetime as a pseudo-Riemannian
  manifold loses sense at Planck scale $$\lambda_P =
  \sqrt{\frac{G\hbar}{c^3}}\simeq 1.6 \times 10^{-33}cm.$$  
This is because an arbitrary accurate localization process requires to
  concentrate an arbitrary amount of energy in a small volume, yielding the
  creation of a black hole. To maintain an operational meaning to the
  measurement process, one should impose some non-zero minimal uncertainties
  in the simultaneous measurement of spacetime coordinates. This  can
  be realized by promoting the coordinates functions $x^\mu$
to operators $q^\mu$ satisfying the non-commutative relation
\begin{equation}
[q_\mu, q_\nu]
= Q_{\mu\nu}.\label{eq:1}
\end{equation}

The behavior of \eqref{eq:1} under a Poincar\'e
transformation marks the difference between two classes of models,
that both have given birth to an extended literature and many
discussions, sometimes quite vivid.
For simplicity, let us assume that the commutator of two coordinates
is a central element (although some models of non-central commutators
have been investigated, see in Tomassini's paper), that is
\begin{equation}
\label{commcentral}
[q_\mu,q_\nu] = i\lambda_P^2 \theta_{\mu\nu}\mathbb I
\end{equation}
where $\I$ is the identity operator in the Hilbert space on which the
$q^\mu$ are represented and $\Theta=\left\{\theta_{\mu\nu}\right\}$ is
an antisymmetric matrix. Obviously \eqref{commcentral} is not invariant under the action of the Poincar\'e group
\begin{equation}
q_\mu\mapsto q'_\mu \doteq \Lambda_\mu^\alpha q_\alpha + a_\mu
\mathbb I
\label{eq:6} \quad \quad \Lambda\in
SO(3,1), a\in \mathbb R^3
\end{equation}
 since 
\medskip
\begin{align}
  \label{eq:13bis}
  [q'_\mu, q'_\nu] &=  
 [a_\mu, a_\nu]\I  +  q_\alpha([a_\mu\I, \Lambda_\nu^\alpha\I] -
 [a_\nu\I,\Lambda_\mu^\alpha\I]) + \Lambda_\mu^\alpha[q_\alpha,
 q_\beta]\Lambda_\nu^\beta \\&=  i\lambda_P^2\,\Lambda_\mu^\alpha
  \Lambda_\nu^\beta \theta_{\alpha\beta}\I \neq    [q'_\mu, q'_\nu].
\end{align}

One may ask that $\Theta$ transforms under the conjugate action of
the Poincar\'e group, yielding 
the \emph{Poincar\'e covariant} model of Doplicher, Fredenhagen,
Roberts.
The Planck length, viewed as the norm of the tensor
$\lambda_P\Theta$ is Poincar\'e invariant, and there is no
modification of the dispersion relation $E^2 = p^2 c^2 + m^2
c^4$. Applications of this model to cosmology are presented in {\bf
  Luca Tomassini}'s \emph{Noncommutative Friedmann spacetimes from
  Penrose-like inequalities} \cite{Tomassini:2015aa}.

\subsection{Deformed-Poincar\'e invariant spacetime}

Alternatively one may require the \emph{Poincar\'e invariance} of the
relation \eqref{commcentral} by  imposing
\begin{equation}
[q'_\mu, q'_\nu] =  i\lambda_P^2
\theta_{\mu\nu}.\label{eq:2}
\end{equation}
This means that the symmetry group of the quantum space is no longer
the Poincar\'e group, but a quantum group deformation of it (the so
called $\theta$-Poincar\'e quantum
group), charac\-terized by a
non-trivial commutation relation for
translations
\begin{equation}
  \label{eq:52}
[a_\mu, a_\nu] = i\theta_{\mu\nu}-
i\theta^{\alpha\beta}\Lambda^\mu_\alpha \Lambda^\nu_\beta.  
\end{equation}
Under this symmetry, $\lambda_P$ is again Lorentz invariant but 
there is now a
  modification of the dispersion relation
  \begin{equation}
E^2 = p^2c^2 +
  m^2 c^4 + f(m,p,E).
\label{eq:4}
  \end{equation}
The possible experimental signature of 
\eqref{eq:4} have been intensively explored, also for the Lie algebra-
like noncommutativity ($\kappa$-Poincar\'e deformation)
\begin{equation}
[q_\mu, q_\nu]
  =\kappa \,\epsilon_{\mu\nu}^\rho q_\rho.
\label{eq:3}
  \end{equation}
Such quantum-group deformations of Poincar\'e symmetries provide 
a useful tool to implement the idea of \emph{Doubly Special
  Relativity}, that is the implementation of the Planck length
$\lambda_P$ as an invariant scale, in a similar  manner as the speed
of light $c$ \cite{Amelino-Camelia:2002jw}. This has been interpreted later as
a geometry where the space of momenta is curved (\emph{Relative
  Locality}). Recent developments on that matter are treated in {\bf
  Stefano Bianco} \emph{Phenomenology from the
  DSR-deformed relativistic symmetries of 3D quantum gravity via the
  relative-locality framework} \cite{Bianco:2015aa}. 

\subsection{Quantum field theory on noncommutative space}

Quantum field theories on noncommutative spacetime were put at the front
of the scene by (open) string theory, when it was observed that $D$-brane world volume acquire
a noncommutative deformation in the background of a non-zero
B-field. This, together with new developments on closed string and
nonassociative algebras, is recalled in {\bf Paolo Aschieri \& Richard J. Szabo}
\emph{Triproducts, nonassociative strar products and geometry of
  R-flux string compactifications} \cite{Aschieri:2015aa}.

\medskip

However one should not forget that quantum field and gauge theories on
noncommutative spacetimes have been originally introduced independently of
quantum gravity, as a tool  to avoid ultraviolet divergencies. This
was the original idea of Snyder, that has been somehow subsumed by
renormalization. Nevertheless the idea that noncommutative spaces
offer a beautiful ground to understand better quantum field and gauge
theories, especially their renormalization properties, has been
intensively investigated in the last decade. A scalar field model on a
generalized $\kappa$-Minkowski space is
presented in {\bf Tajron Juric, Stjepan Meljanac, and Andjelo
  Samsarov} \emph{Light-like $\kappa$-deformation and scalar field
      theory via Drinfeld twist} \cite{T.-Juric:2015aa}. The recent advances in gauge theory on
    noncommutative spacetimes \cite{mart-matrix} are reported in {\bf Jean-Christophe
      Wallet} \emph{Spectral theorem in noncommutative field theories:
      Jacobi dynamics} \cite{Gere:2014fk}.
\medskip

Finally, to go back to quantum gravity, let us also mention that Snyder's ideas
re-thought from a quantum gravity perspective yields intriguing
results. Recently, it has been used to question how much of the noncommutativity of the
coordinates actually survive the description of physical systems. This is the object of {\bf Valerio Astuti}
\emph{Covariant quantum mechanics applied to noncommutative geometry} \cite{Astuti:2015aa}.

\newpage \section{Spectral geometry}

On may question the physical meaning of the noncommutative coordinate
$q^\mu$ in \eqref{eq:1}. As elements of the (abstract) polynomial algebra they
generate, the spectrum of the $q^\mu$'s is not real, which makes their
interpretation as physical observable problematic. A solution is to  represent the $q^\mu$'s  on
some Hilbert space, as this is done in quantum mechanics. To do so, it is convenient to
view the $q_\mu$'s as affiliated to the algebra of compact
operators, as pointed out in
the Doplicher-Fredenhagen-Roberts. But this indicates that the algebra generated by the coordinates may not
be the most accurate tools to describe a quantum space, an
algebra suitably associated to the $q^\mu$'s can do a better job. 
With this idea in minds, one has no reason to restrict one's attention
to noncommutative deformations of
  commutative coordinates: there are much more noncommutative algebras
  to play with ! This idea is enforced by Gelfand duality between 
 commutative $C^*$-algebras and locally compact spaces, which suggests
 that a natural definition of a  \emph{noncommutative
geometry} is an object such that its ``algebra of functions'' (and not
only its coordinates) is noncommutative.
 
\subsection{The standard model and beyond}
  Connes' theory of spectral triples $(\A, \HH, D)$  extends
  Gelfand duality,
 beyond
 topology, so that to encompass all the geometrical aspects of
 Riemannian geometry. 

A spectral triple consists in a involutive algebra $\A$, 
a faithful representation on $\HH$, a densely defined operator $D$ on $\HH$ such
that $[D, a]$ is bounded and  $\; a[D - \lambda\I]^{-1}$ is compact
for any $a\in \A$ and $\lambda\notin \text{ Sp } D$. 
Imposing a set of further conditions, one defines \emph{real} spectral
 triples,  whose paradigmatic example is
 \begin{equation}
 (C^\infty\!(\M), L^2(\M, S), \slash\!\!\! \partial  =
 -i\gamma^\mu\partial_\mu),
\label{eq:7}
 \end{equation}
that is the algebra of smooth functions on a closed spin manifold
$\M$, acting on the Hilbert space of square integrable spinors, with
$D=\slash\!\!\! \partial$ the Dirac operator. Conversely, one has the
following reconstruction theorem \cite{connesreconstruct}:
given a real spectral triple $(\A, \HH, D)$  with $\A$ unital commutative,  then
 there exists a compact oriented Riemannian spin manifold $\M$ such that
 $\A=C^\infty(\M)$.
These conditions still makes sense for non-commutative $\A$, so that
one defines a \emph{noncommutative geometry} as a spectral triple $(\A, {\cal
      H}, D)$ where $\A$ is non necessarily commutative. To summarize:
\begin{eqnarray*}
 \text{commutative spectral triple} &\rightarrow& \text{noncommutative
   spectral triple}\\
 \updownarrow & & \downarrow \\
 \text{Riemannian geometry} & & \text{non-commutative geometry}
 \end{eqnarray*}

Spectral triple turn out to be a powerful tool to describe the
standard model of elementary particles from a purely geometric point
of view. The starting point is to view spacetime no more as a
manifold,  as in general
relativity, but as the product of a manifold by a matrix
geometry. This allows to incorporate in the geometry the degrees of
freedom of the standard model. More precisely, one
considers the  \emph{almost-commutative} algebra
\begin{equation}
  \label{eq:1bis}
  \cinf \otimes \A_F,
\end{equation}
where $\A_F$ is a finite dimensional algebra that carries the gauge
group of the standard model (which is retrieved as the group of
unitaries of $\A_F$). It acts on the space of fermions, that is ${\mathbb C}^{96}$
(two colored quarks, one neutrino and one electron make $8$, that
multiplies $2$ for the chirality, another $2$ for antiparticles and $3$
for the number of generations). There is a finite dimensional Dirac
operator $D_F$, that is a $96\times 96$ matrix whose entries are the
Yukawa coupling of the fermions and the mixing angles of quarks and
neutrinos. A general formula for product of spectral triples yields the
generalized Dirac operator $D:=\slash\!\!\! \partial\otimes \I_{96} +
\gamma^5\otimes D_F$. Bosonic fields are generated by the so-called fluctuations of the metric, that is the substitution of $D$  with
the covariant Dirac operator
\begin{equation}
  \label{eq:8}
  D_A:= D + A + J AJ^{-1}
\end{equation}
where $A$ is a selfadjoint element of the set of generalized $1$-forms
$\Omega^1_D:=\left\{ a^i [D, b_i]\right\}$. 
The asymptotic expansion $\Lambda\to\infty$ of the \emph{spectral
  action} $\text{Tr} f(\frac D{\Lambda})$ \cite{Chamseddine:1996kx} where
$\Lambda$ is a cutoff parameter and $f$ a smooth approximation of the
characteristic function of the interval $[0,1]$ yields the bosonic
Lagrangian of the standard model (including the Higgs)
minimally coupled to 
Einstein-Hilbert action (in Euclidean signature).

In other terms the standard model appears as a purely gravitational theory, but on a (slightly)
noncommutative space. 
As a bonus, the Higgs field comes out as the component of a connection in
  the noncommutative part of the geometry. 

Practically, the spectral action provides relations between the parameters of the
theory at a putative energy of unification. In particular the mass
term of the Higgs appears as a function of the input of the models,
namely the Yukawa couplings of fermions.  Assuming the big-desert
hypothesis, the running of this mass under the flow of the renormalization group
yields a prediction for the mass of the Higgs of $170$ GeV, a value
ruled out by Tevatron in $2008$.   

Since then the Higgs-Brout-Englert boson has been discovered with a mass around 
$125$ GeV. This mass is problematic, or at least intriguing, because
it lies just below the threshold of stability, meaning that
electroweak vacuum is a metastable state rather than a stable one.
One solution to stabilize the electroweak vacuum is to postulate there
exists another scalar field, called $\sigma$, suitably coupled to the
Higgs. Chamseddine and Connes have noticed in \cite{Chamseddine:2012fk} that taking into account this
new scalar field in the spectral action, by promoting the Yukawa
coupling of the right neutrino (which is one of the constant component
of the matrix $D_F$) to a field,
\begin{equation}
  \label{eq:5}
  k_R \to k_R\sigma,
\end{equation}
 then one obtains the correct coupling to the Higgs as well as a way to pull back the mass of the Higgs from $170$ to
  $126 \text{GeV}$. In \cite{Devastato:2013fk,buckley}, as reported in
  {\bf Agostino Devastato} \emph{Noncommutative geometry, grand
    symmetry and twisted spectral triple} \cite{Devastato:2015aa}, its is 
  shown how the substitution \eqref{eq:5} can be obtained as a
  fluctuation of the Dirac operator, but in a slightly modified
  version inspired by the notion of \emph{twisted spectral triple}
  introduced previously by Connes and Moscovici. The field $\sigma$
  thus appears as a Higgs-like field associated to a spontaneous
  symmetry breaking to the standard model of a ``grand symmetry'' model where the spin degrees of
 freedom ($C^\infty({\cal M})$ acting on the space of spinors) are mixed with the
 internal degrees of freedom ($\A_F$ acting on the space of
 particles).


\subsection{Noncommutative space versus noncommutative space-time}

Spectral triple provides a generalization of Riemannian geometry to
the noncommutative setting, but there is no reconstruction theorem for
pseudo-Riemannnian manifolds. Once computed the spectral
action, one makes a Wick rotation $t\to i t$, as done for
instance in the path integral approach to quantum gravity. However one
might like to make sense of
Minkovskian signature from the beginning. In \emph{Two roads to
  noncommutative causality} \cite{Besnard:2015aa}, 
{\bf Fabien Besnard} presents a state of the art, including his own recent
results, on various attempts to incorporate a causal structure in
spectral triples.
\medskip

Let us end this discussion by our own contribution, stressing the
algebraic approach on how to put time into the game. This
is the idea, sometimes advertised by Connes as ``the heart of noncommutative geometry'',
that time involution is intrinsically contained within the noncommutativity of the algebra. Namely, given a von Neumann algebra $\A$ acting on an Hilbert space $\cal H$ together with a 
 vector $\Omega$ in $\cal H$ cyclic and separating for the action of $\A$, one defines by Tomita-Takesaki modular theory a $1$-parameter group of
 automorphism $\sigma^\Omega_t\in\text{Aut}(\A)$. Connes has shown
 that the group obtained from another state $\Omega'$ differs from the
 former only by unitaries, that is
 \begin{equation}
   \label{eq:9}
   \sigma^{\Omega'}_t = U^{\Omega'\Omega}_t \sigma^\Omega (
   U^{\Omega'\Omega}_t)^* \quad \forall t\in \mathbb R
 \end{equation}
where the unitary intertwining is given by Connes cocycle
$U^{\Omega'\Omega}_t$.
Hence there is a canonical group of outer automorphism $\sigma_t$ canonically
associated to the von Neumann algebra $\A$, where
$$\text{Out} (\A({\cal O})) =\text{Aut} (\A({\cal O}))\slash \text{Inn}
 (\A({\cal O})).$$

 The physical
interpretation of this group as a time evolution is enforced by the
fact that $\sigma_\Omega^t$ satisfies with respect to
$\Omega$ the same properties as does in statistical physics the time
evolution $e^{iHt}\cdot e^{-iHt}$ with respect to a Gibbs equilibrium state,
namely the KMS condition. 

\begin{figure}\centering
 \mbox{\rotatebox{0}{\scalebox{.5}{\includegraphics{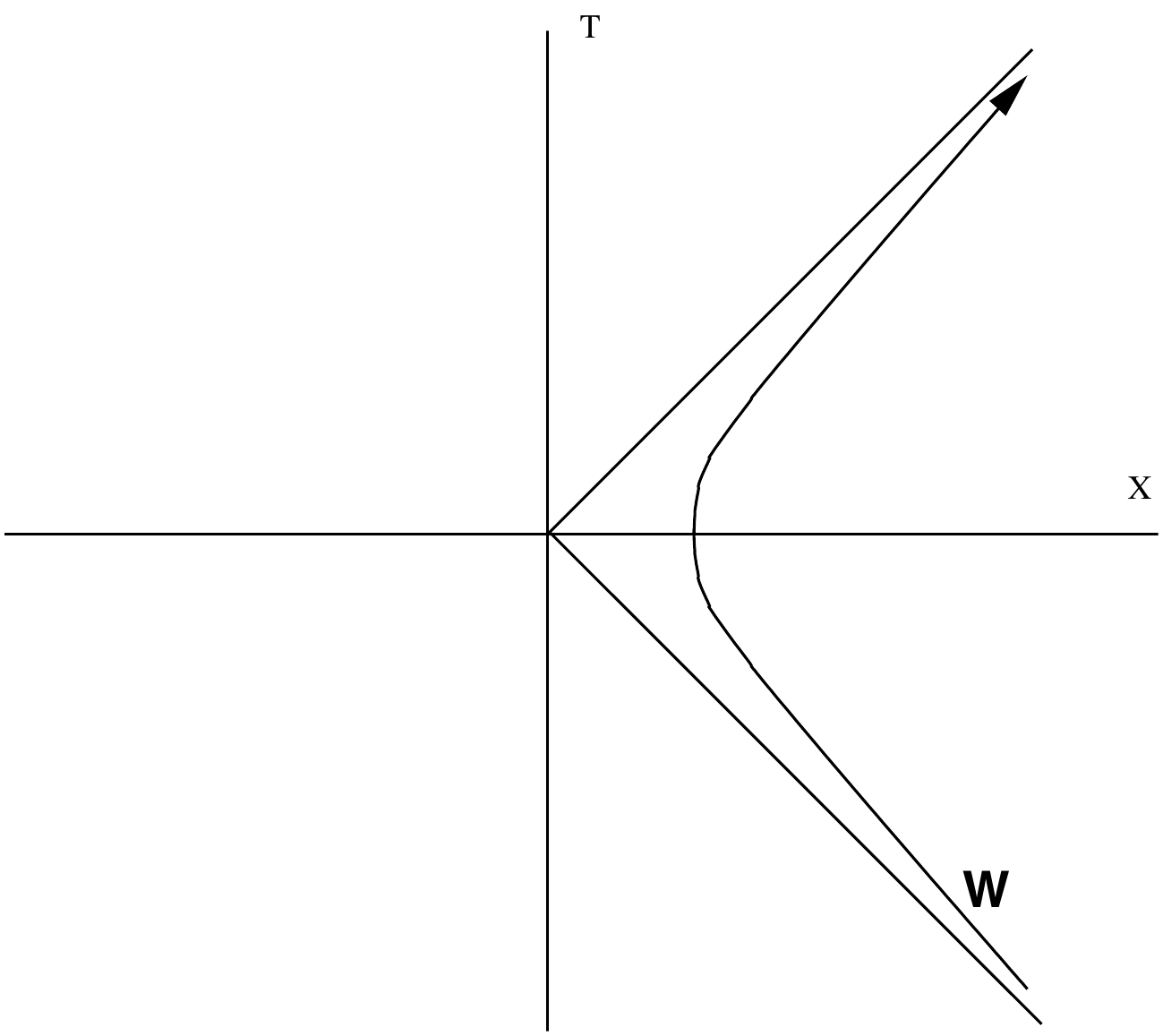}}}}
 \hspace{1truecm}
 \mbox{\rotatebox{0}{\scalebox{.5}{\includegraphics{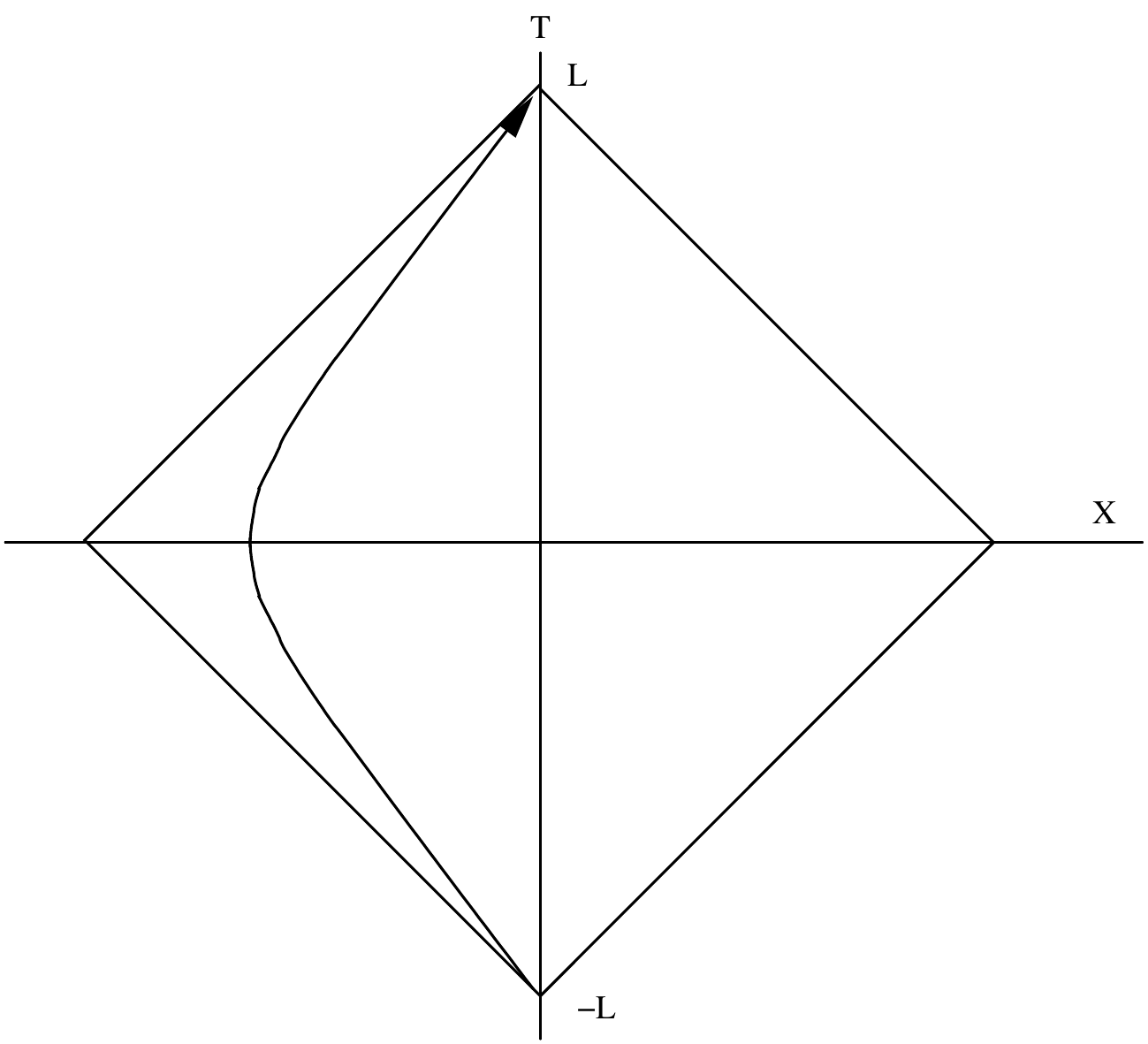}}}}
\caption{\label{fig1}An orbit of the vacuum modular group for the algebra
  of local observables localized in the Rindler wedge (left) and in a
  double-cone of Minkowski spacetime (right).}
\end{figure}

An interesting framework to test the interpretation of the
modular group as a real "physical" time is algebraic quantum field theory, where the von Neumann
algebra is the algebra $\A({\mathcal O})$ of local observables
associated to an open region ${\mathcal O}$ of Minkowski spacetime. In
particular, for $\mathcal O$ the Rindler Wedge, it is well known that the
modular group of the vacuum vector is generated by the boosts. Hence
it has a geometrical action whose orbits are the trajectories of
uniformly accelerated observers. The KMS condition is interpreted
as the vacuum being a thermal equilibrium state for this
observer, with a temperature proportional to the acceleration (see
\cite{Buchholz:2014aa} for a recent critical view on this
interpretation, though). A similar analysis holds for double-cone regions of
Minkowski spacetime and yields a correction to the Unruh temperature
for an observer with a finite life-time \cite{Martinetti:2003sp}, see
fig. \ref{fig1}.

\begin{figure}
 \centering
\vspace{-3.5truecm}\mbox{\rotatebox{0}{\scalebox{.5}{\includegraphics{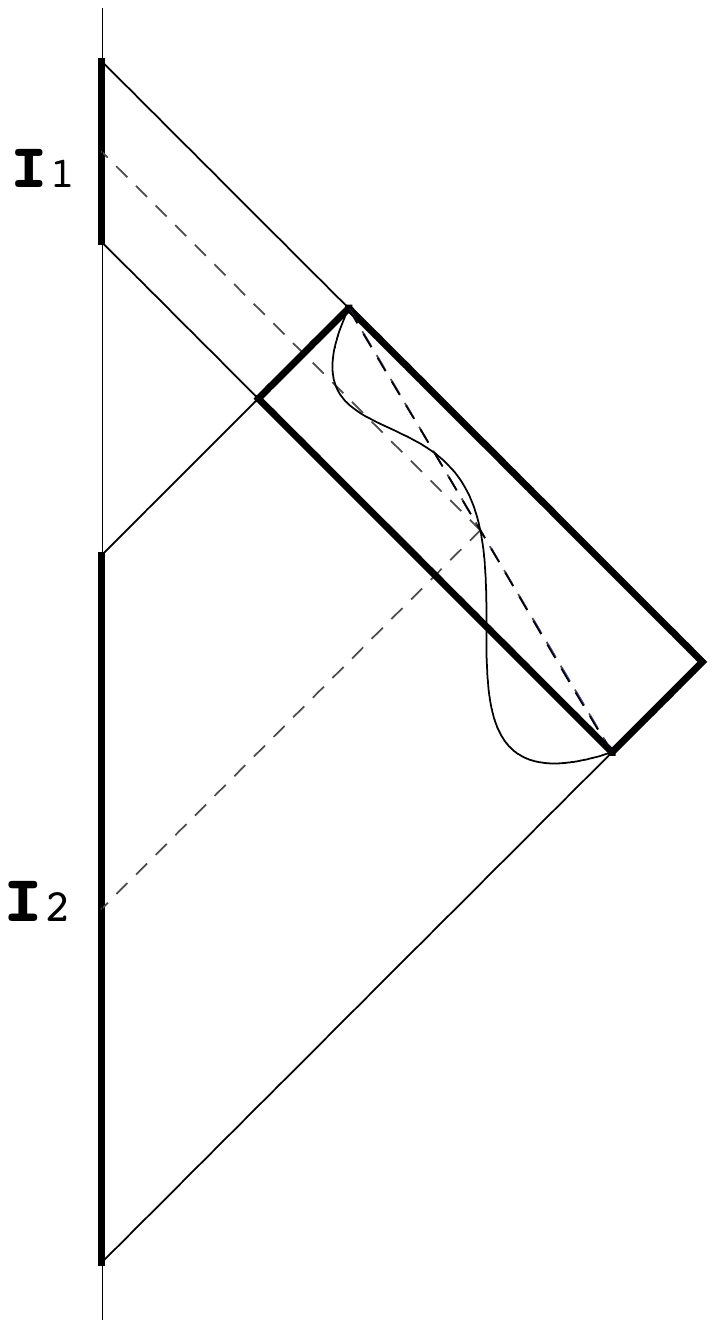}}}}
\vspace{-3truecm} 
\caption{\label{fig2} A modular orbit in $2D$-conformal theory with
  boundary.}\end{figure}

 Interestingly, a similar
analysis for a double-cone in a bidimensional boundary conformal field theory
permits to compute explicitly the action of Connes cocycle. The field in the
    double-cone is determined by its components $\psi(x_1),
    \psi(x_2)$ on each interval on the boundary defining the
    double-cone (fig. \ref{fig2}). There is a state whose modular group
    has pure geometrical action $x_1(t), x_2(t)$ (but the orbit is not
    the trajectory of an observer with constant acceleration) while the
modular group for the vacuum mixes this geometric action with a mixing
of the components
$$\hspace{-1truecm}\sqrt{\frac{d x_i}{d\zeta}} \,\sigma_t(\psi(x_i)) =
\underset{k = 1,2}{\sum}
O_{ik}(t) \sqrt{\frac{d x_k}{d\zeta}}\,\psi(x_k(t))$$
where $\zeta$ is a suitable parametrization of the orbit and $O_{ik}$
is the mixing matrix. In this
sense, the action of the unitary cocyle is non-geometric, and amounts
to mix the components of the conformal field on the two intervals. For
a further interpretation of this, see \cite{Rehren:2012fk} and \cite{Martinetti:2012fk}.

Besides quantum field theory, the hope is that
this way of extracting a time flow from an algebra of observables
and a state may be relevant in quantum gravity \cite{Connes:1994xy}.

\end{document}